\documentclass[a4paper,12pt]{article}
\pdfoutput=1
\usepackage[T2A]{fontenc}
\usepackage[utf8]{inputenc}
\usepackage[english]{babel}
\usepackage{lineno}
\modulolinenumbers[5]

\usepackage[warn]{mathtext}	
\usepackage{graphicx} 
\usepackage{indentfirst}
\usepackage{psfrag}

\usepackage{geometry}
\makeatletter
\renewcommand{\@biblabel}[1]{#1.}
\makeatother

\usepackage{amssymb,amsfonts,amsmath,mathtext,enumerate,float}
\usepackage{indentfirst}	
\usepackage{color,soul}

\begin{document}
\begin{center}
{\Large\textbf{Experimental studies of thorium ions implantation from pulse laser plasma into thin silicon oxide layers}}\\

P.V.~Borisyuk$^{1,*}$, E.V.~Chubunova$^1$, Yu.Yu.~Lebedinskii$^{1,2}$, E.V.~Tkalya$^{1,3,4}$, O.S. Vasilyev$^1$, V.P. Yakovlev$^1$, E.~Strugovshchikov$^{1,5}$, D.~Mamedov$^1$, A.~Pishtshev$^6$, S.Zh.~Karazhanov$^{1,5}$

\end{center}

{\small{\noindent$^1${National Research Nuclear University MEPhI (Moscow Engineering Physics Institute), 31 Kashirskoe sh., Moscow, 115409 Russian Federation}\\
$^2${Moscow Institute of Physics and Technology (State University), 9 Institutskiy per., Dolgoprudny, Moscow Region, 141700, Russian Federation}\\
$^3${Skobeltsyn Institute of Nuclear Physics, Lomonosov Moscow State University, Leninskie gory, Moscow 119991, Russian Federation}\\
$^4${Nuclear Safety Institute of RAS, Bol'shaya Tulskaya 52, Moscow 115191, Russian Federation}\\
$^5${Department for Solar energy, Institute for Energy Technology, 2027 Kjeller, Norway}\\
$^6${Institute of Physics, University of Tartu, 50411 Tartu, Estonia}\\
$^*$ PVBorisyuk@mephi.ru}}\\

\textbf{Abstract}
{\small{We report the results of experimental studies related to implantation of thorium ions into thin silicon dioxide by pulsed plasma fluxes expansion. {Thorium ions were generated by laser ablation from a metal target, and the ionic component of the laser plasma was accelerated in an electric field created by the potential difference  (5, 10 and 15 kV) between the ablated target and SiO$_2$/Si (001) sample. Laser ablation system installed inside the vacuum chamber of the electron spectrometer was equipped with YAG:Nd3+ laser having the pulse energy of 100 mJ and time duration of 15 ns in the Q-switched regime. Depth profile of thorium atoms implanted into the 10 nm thick subsurface areas together with their chemical state as well as } the band gap of the modified silicon oxide at different conditions of implantation processes were studied by means of X-ray photoelectron spectroscopy (XPS) and Reflected Electron Energy Loss Spectroscopy (REELS) methods. Analysis of chemical composition showed that the modified silicon oxide film contains complex thorium silicates. Depending on local concentration of thorium atoms, the experimentally established band gaps were {located} in the range of 6.0~--~9.0~eV. Theoretical studies of optical properties of the SiO${}_{2}$ and ThO${}_{2}$ crystalline systems have been performed by ab initio calculations within hybrid functional. Optical properties of the SiO${}_{2}$/ThO${}_{2}$ composite were interpreted {on the basis of} Bruggeman effective medium approximation. A quantitative assessment of the yield of isomeric nuclei in "hot" laser plasma at the early stages of expansion has been performed. The estimates made with experimental results demonstrated that the laser implantation of thorium ions into the SiO${}_{2}$ matrix can be useful for further research of low-lying isomeric transitions in ${}^{229}$Th isotope with energy of 7.8~$\pm$~0.5~eV.}}\\

Keywords: Wide-band gap compound; Thorium-229; Laser implantation; XPS; REELS; Low lying isomer state.\\

\section{Introduction}
Currently, wide-band gap materials doped with thorium ions are subject of intensive studies \cite{1,2,3,4,5} This is due to the fact that such materials could serve as the most promising for the study of a unique anomalously low-lying isomeric state of the ${}^{229}$Th nucleus \cite{6} with the energy of 7.8 $\pm$ 0.5~eV \cite{7,8}. However, the measurement uncertainty of  $\sim$ 0.5~eV requires a frequency tuning in the range of  $\sim$~250~THz that leads to significant impediment to use precision of laser spectroscopy techniques for direct detection of the desired transition. This circumstance, also, is a limiting factor for developing novel devices operating in the vacuum ultraviolet (VUV) range such as, e.g., a laser on a nuclear transition \cite{9}, time and frequency references of a new generation \cite{10}, etc. Today, the available synchrotron sources may excite the direct energy transitions in the ${}^{229}$Th nuclei which are accommodated in crystals transparent in the VUV region. Possible candidates are  LiCaAlF${}_{6}$, LiSrAlF${}_{6}$, and CaF${}_{2}$ \cite{11,12}. However, synthesis of high purity material containing the required amount of the thorium isotope presents a complex and resource-intensive problem \cite{13,14}. The poor availability of a synchrotron radiation source of high intensity with possibility of tuning the lasing energy in the range 7.2~--~8.3~eV, also makes it difficult to conduct experiments of this kind.

In this paper we propose an original physical system that allows excitations of the ${}^{229}$Th nuclei within a wide energy range with significant increase in the fluorescent signal produced as a result of the relaxation of the populated isomeric state. Experimentally, the system is formed by means of the pulsed laser implantation; it consists of  an ensemble of the ${}^{229}$Th ions embedded into the matrix of a broadband dielectric medium (silicon oxide) by means of the pulsed laser implantation. A key feature of the system is that the thorium ions are subject to a large number of inelastic collisions with the neighboring electrons at the early stages of the expansion of the plasma torch when plasma can still be treated as quasi-stationary. Due to the mechanism of Inverse Internal Electronic Conversion (IIC) \cite{15}, this process can move the ${}^{229}$Th nucleus into an excited state. Thus, laser implantation offers a unique opportunity allowing us to combine two important processes: 1~--~excitation of the ${}^{229}$Th isomeric nuclei; 2~--~implantation of a wideband dielectric into the matrix. At the same time, the key issue of this work is the study of physicochemical properties of thorium-containing samples obtained during laser implantation. It is important that the band gap should be larger than 7.8 eV, and the number of implanted nuclei should be sufficient to register a VUV photon.

We present an analysis of the electronic structure of thin thorium-containing silicon oxide layers formed by pulsed laser implantation. Using the X-ray Photoelectron Spectroscopy (XPS) and Reflected Electron Energy Loss Spectroscopy (REELS), the chemical environment and the band gap of the formed layers for a different number of implanted thorium ions were extracted. The measured value of the band gap is compared with the calculated value obtained from the first principles. The possibility of using such a system for further studies of a nuclear low-lying isomeric transition in the ${}^{229}$Th is under discussion.

\section{Experimental setup}

Experimental studies were conducted using an XSAM-800 electronic spectrometer (Kratos) \textit{in situ} equipped with pulsed laser deposition (PLD) \cite{16}. Production of ions, their acceleration in an external electric field followed by the process of implantation are routine applications of PLD method actively utilized in Surface Science \cite{17,18,39}. {In particular some details of the PLD technique one can find in Ref.~\cite{40}} In our experiments thorium was evaporated by laser ablation from a metal target, and the ionic component of the laser plasma was accelerated in an electric field {created} by the potential difference (5, 10 and 15~kV) between the ablated target and SiO${}_{2}$/Si~(001)  sample. {As a substrate for thorium implantation we used a thermally oxidized and highly doped (0.01 Ohm$\cdot$cm) silicon wafer Si (001) with a 600 nm thick SiO$_2$  film. The PLD depositions were performed on the freshly cleaned substrate surface SiO$_2$/Si(001) under ultra high vacuum ($5\times10^{-9}$ Torr) conditions  at room temperature. The PLD system was equipped with YAG:Nd3+ laser (E = 100 mJ) and installed inside the vacuum chamber of the electron spectrometer. Deposition rate can be varied from $10^{13}$ to $10^{15}$ cm$^{-2}$ per pulse with time duration of 15 ns in the Q-switched regime.} 

Immediately after each series of pulses of thorium sputter in an electric field, the Th/SiO${}_{2}$/Si~(001) sample was transferred into the spectrometer analyzer chamber under the high vacuum where XPS and REELS analyses were performed. To excite photoelectrons we used an X-ray radiation of the AlK$\alpha$${}_{1,2}$ line with an energy of \textit{h$\nu$}~=~1486.6~eV. The spectrometer was calibrated along the Au4f${}_{7/2}$ line with the binding energy BE~=~84.0~eV.

The band gap width of the substrate and the layers formed on its surface were measured by the REELS method. The electron beam energy was E${}_{0}$~=~1000~eV, the beam current was I${}_{0}~\approx 30$~$\mu$A, the scattering angle was equal to $\varphi~{}_{0}$~=~$125^{\circ}$$~\pm~ 20^{\circ}$ (backscattered electrons were collected within the angle $\Delta~\varphi$~=~$\pm~20^{\circ}$). The energy width of the primary beam electrons was $\Delta$~E~$\approx$~1.5~eV after reflection.

We note that the study was carried out using a metal target of the ${}^{232}$Th natural isotope due to the high radioactivity of the ${}^{229}$Th isotope. This enabled us to reduce the level of radioactivity of the samples by the level allowed to conduct the necessary series of experiments.

\section{Experimental results}

XPS spectra of Si2p and {Th4f$_{7/2,5/2}$ lines for samples with different Th/Si atomic ratio inside the subsurface area formed as a result of Th implantation process via PLD depositions under the voltage as high as 15 kV are presented in Fig.~1. For comparison the lowest curve on the left hand side picture shows the Si2p line for the initially pure SiO$_2$ surface just before the start of implantation process. Standard spin–orbit splitting results in the Th4f$_{7/2,5/2}$ doublet-line structure shown on the right hand side picture.} The average thorium concentration in the subsurface area determined by the mean free path ($\sim$~3~nm) of electrons with kinetic energy  $\sim$~1~keV was {extracted from} the intensities of the Si2p and Th4f lines. It can be seen from the figure that the thorium concentration in the subsurface area concerned increases as the number of laser pulses increases, while the Th4f-line position does not change. The Th4f${}_{7/2}$ line has a binding energy BE~=~334.9~eV at all stages of deposition. According to \cite{19} the binding energy of the Th4f${}_{7/2}$ level is equal to 335.0~eV and 335.3~eV for compounds ThSiO${}_{4}$ и ThSiO${}_{4}$ $\cdot$ nH${}_{2}$O, respectively. Since these values are  close to the observed binding energy it can be assumed that a thorium chemical bond is formed as a result of implantation which is close to that of thorium for ThSiO${}_{4}$ in ionicity. In this case the Si2p-line broadens with increasing the Th concentration and shifts toward lower binding energies: from 103.4~eV for SiO${}_{2}$ to 102.6~eV for a film with a ratio Th/Si~=~1. This change in the binding energy of the Si2p line is usually connected with the formation of silicates. In our case it can be due to the formation of a ThSi${}_{x}$O${}_{y}$/SiO${}_{2}$/Si -- type film structure with an amorphous ThSi${}_{x}$O${}_{y}$ film of variable composition \cite{20}. A significant broadening of the Si2p-line may also indicate that thorium silicate as a variable composition presents  in the subsurface area.

\begin{figure}
	\centering
	\includegraphics[width=0.7\linewidth]{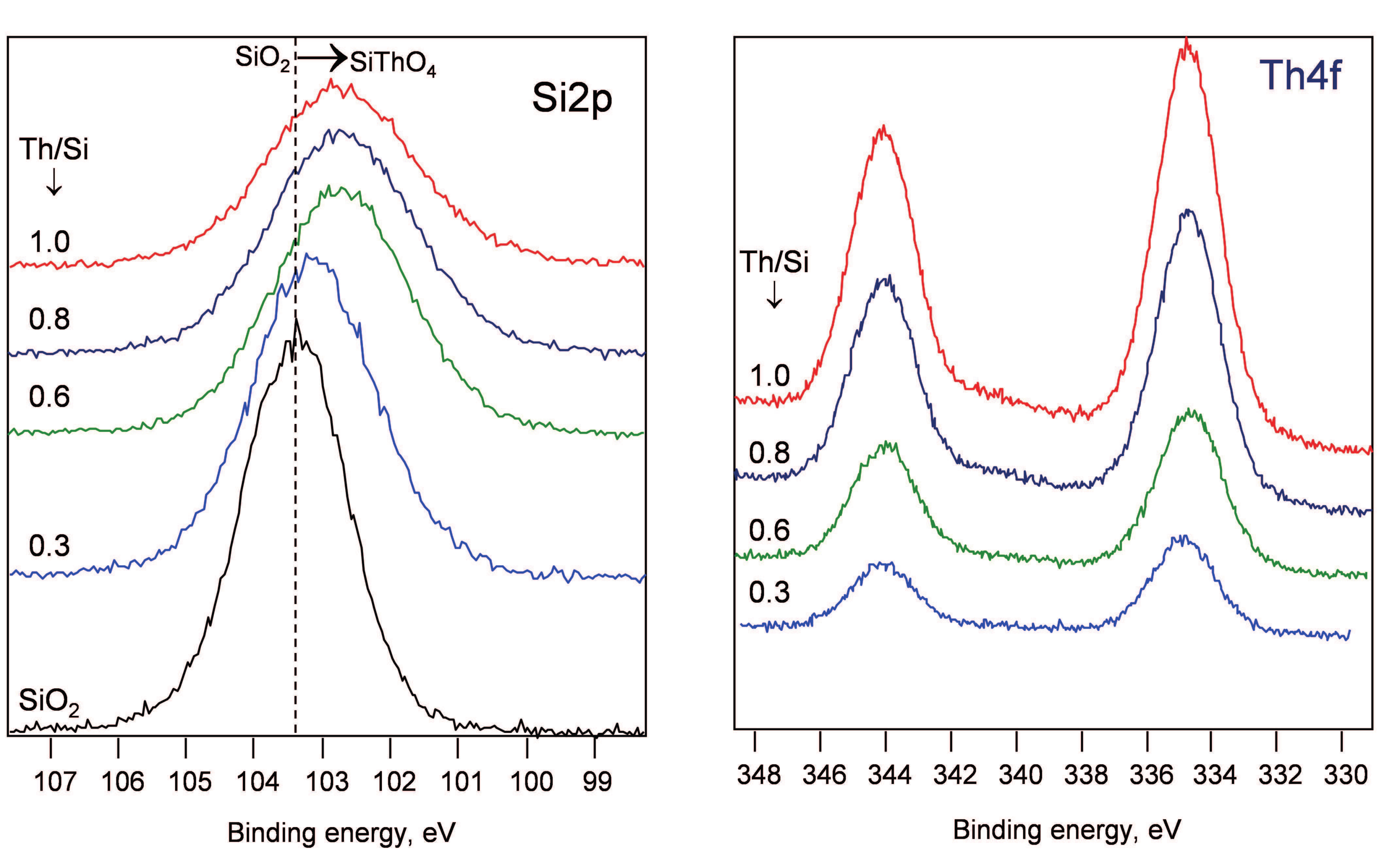}
	\caption{XPS spectra of Si2p and Th4f lines for sample with different Th/Si atomic ratio in near-surface SiTh${}_{x}$O${}_{y}$ region formed after Th implantation during PLD deposition under high voltage 15 kV}
	\label{fig:image1}
\end{figure}

\begin{figure}
	\centering
	\includegraphics[width=0.7\linewidth]{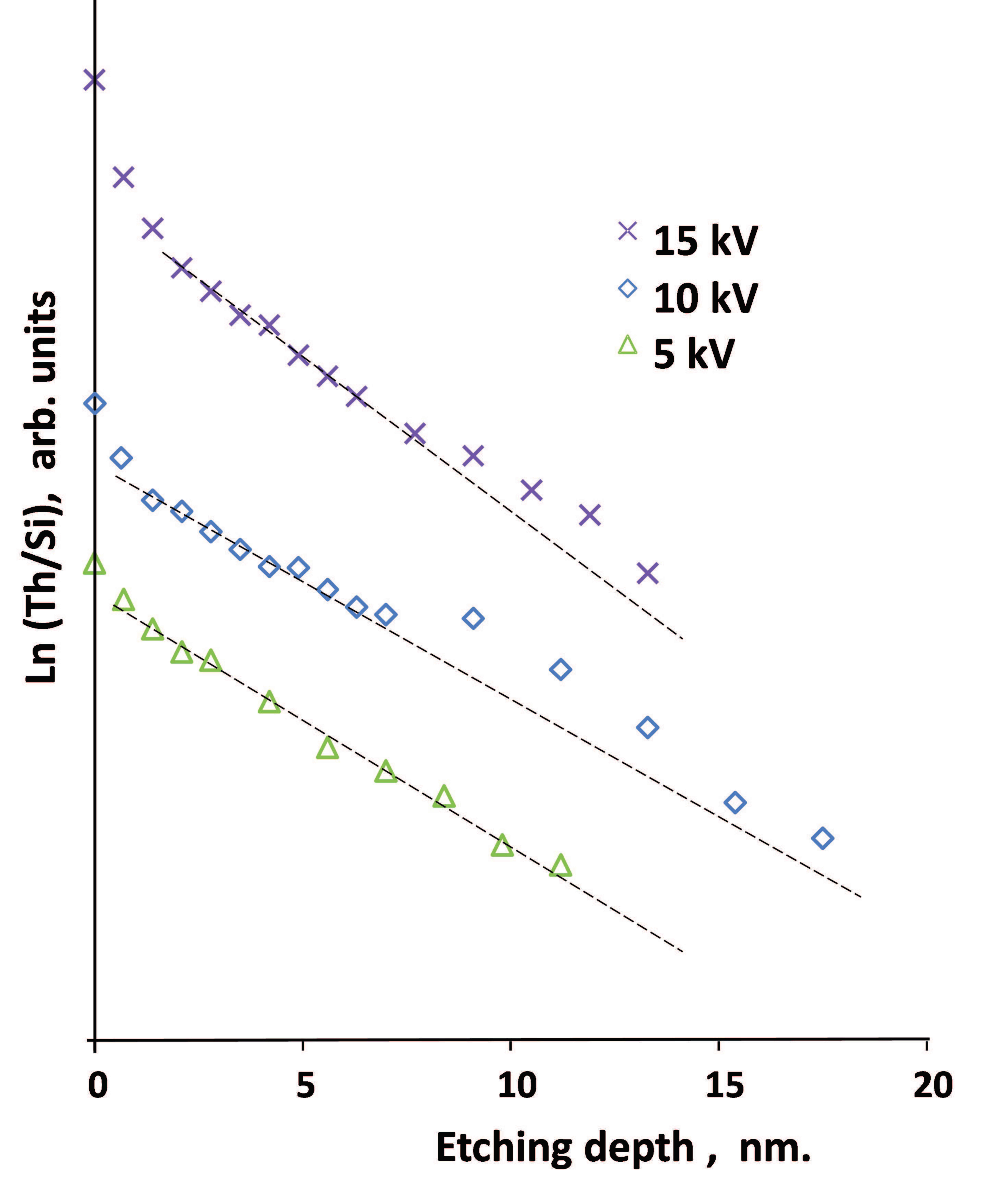}
	\caption{Depth profile with Ar etching for samples ThSiO${}_{X}$/SiO${}_{2}$/Si formed after Th implantation during  PLD deposition under various high voltage}
	\label{fig:image2}
\end{figure}

To obtain depth distribution of implanted thorium atoms, the ion etching XPS method was used in films formed during the deposition with accelerating voltages of 5, 10 and 15~kV. For this purpose the top layer was removed by etching the surface at the rate $\sim$~10~{\AA/min} with Ar$^+$ ions having energy of 3.5~keV. Fig.~\ref{fig:image2} shows logarithmic scale dependences of the change in the ratio of Th/Si intensities on the thickness of the sputtered layer. As can be seen in the figure the Th/Si ratio decreases exponentially. At the same time the distribution profile of thorium atoms over depth is rather blurred and does not have a clear boundary. Nevertheless, nonmonotonic deviations from the exponential law toward a higher Th concentration is observed in case of dependences obtained at accelerating voltages of 10 and 15~kV at a depth of 10~--~15~nm. This indicates a typical maximum at a depth determined by the energy of implanted particles. It should be noted that the SRIM package calculation \cite{21} predicts the junction of 10~keV thorium atoms in the SiO${}_{2}$ matrix at the level of 10~nm that agrees with features experimentally observed in Fig.~\ref{fig:image2}.

We guess the exponentially decreasing distribution profile along the etching depth could be related to the two factors. The first one is of technical nature, and it is due to the fact that the depth of the XPS analysis ($\sim$~3~nm) is comparable in magnitude to the implantation depth in the subsurface layers ($\sim$~10~nm). Consequently, while profiling in the process of etching with argon ions and multiple atomic surface resputtering, the XPS method decreases the possibility of obtaining a clear profile boundary. The second factor is related to the method of obtaining ions and thorium atoms during laser ablation and needs to be explained in a more detail. In fact, there are both neutral and charged components in the flow of thorium particles formed in plasma under laser ablation. Due to the Coulomb interaction the ionic component in the late stages of the plasma expansion has a rather wide spectrum of energy distribution described by the shifted Maxwell-Boltzmann-Coulomb distribution \cite{22}. {For completeness we mention the paper \cite{41} specially devoted to detail investigations  of thorium ion energy spectra in laser plasma.} The Coulomb repulsion and the effect of motion of a plasma bundle as a whole can lead to an energy broadening of the ion beam (W~$\sim$~1~keV) and to blurring the boundary of thorium ions with energy (\textit{E${}_{0}$}~=~10~keV) implanted to a depth of \textit{h}~=~10~nm at a level of several nanometers \textit{$\delta$h}~$\sim$~ \textit{h}$\times $\textit{W}/\textit{E${}_{0}$}. In Fig.~\ref{fig:image2}, a peculiar manifestation of non-monotonicity occurs in the depth interval of 10 to 15~nm, therefore, the blurring of the implantation depth does not exceed 3~nm, that is consistent with the estimation. Hence, the presence of an exponential decay can not be explained by the broad energy distribution of ions. Apparently, the neutral component, being accelerated only in the early stages of the plasma spread when the particle concentration is close to a solid one, gives rise to a beam of a large number of low-energy particles. This implies that the flux of neutral thorium particles entering the surface of SiO${}_{2}$ that have no sufficient energy to implant is deposited into the upper layer. In addition, one should take into account the effect of droplets which is inevitable in case of pulsed laser deposition The Coulomb repulsion and the effect of motion of a plasma bundle as a whole can lead to an energy broadening of the ion beam (W~$\sim$~1~keV) and to blurring the boundary of thorium ions with energy (\textit{E${}_{0}$}~=~10~keV) implanted to a depth of \textit{h}~=~10 nm at a level of several nanometers \textit{$\delta $h} $\sim$ \textit{h}$\times $\textit{W}/\textit{E${}_{0}$}. In Fig.~\ref{fig:image2}, a peculiar manifestation of non-monotonicity occurs in the depth interval of~10 to~15~nm, therefore, the blurring of the implantation depth does not exceed 3~nm, that is consistent with the estimation. Hence, the presence of an exponential decay can not be explained by the broad energy distribution of ions. Apparently, the neutral component, being accelerated only in the early stages of the plasma spread when the particle concentration is close to a solid one, gives rise to a beam of a large number of low-energy particles. This implies that neutral thorium particles arriving at the SiO${}_{2}$ surface have no sufficient energy to be implanted and therefore are deposited into the upper layer. In addition, one should take into account the effect of droplets and surfaces damage which is inevitable in case of pulsed laser deposition \cite{22,23}. While analyzing the energy band gap of the test samples by the REELS method, such effects can make significant changes. However, when the number of laser pulses is small and the concentration of the deposited thorium atoms does not exceed the concentration of substrate atoms, the influence of surface thorium atoms in the scale range of the depth of the REELS analysis ($\sim$~3~nm) can be ignored.

\begin{figure}
	\centering
	\includegraphics[width=0.7\linewidth]{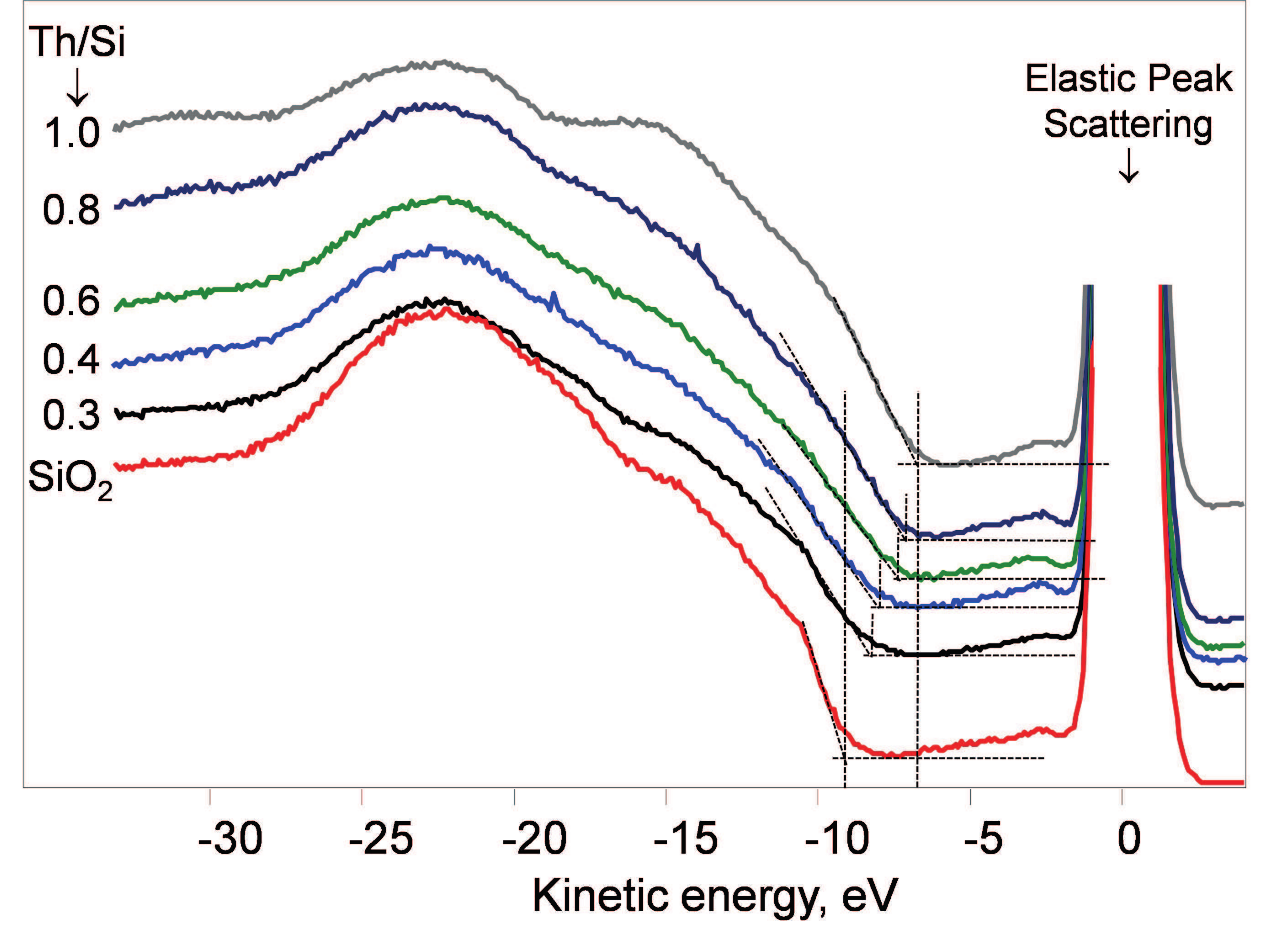}
	\caption{Reflection Electron Energy-loss Spectra with energy 1000~eV scattered on surface of Si/SiO2 samples for different Th/Si atomic ratio in near-surface ThSi${}_{x}$O${}_{y}$ region formed after Th implantation during  PLD deposition under high voltage 15~kV}
	\label{fig:image3}
\end{figure}

\begin{figure}
	\centering
	\includegraphics[width=0.7\linewidth]{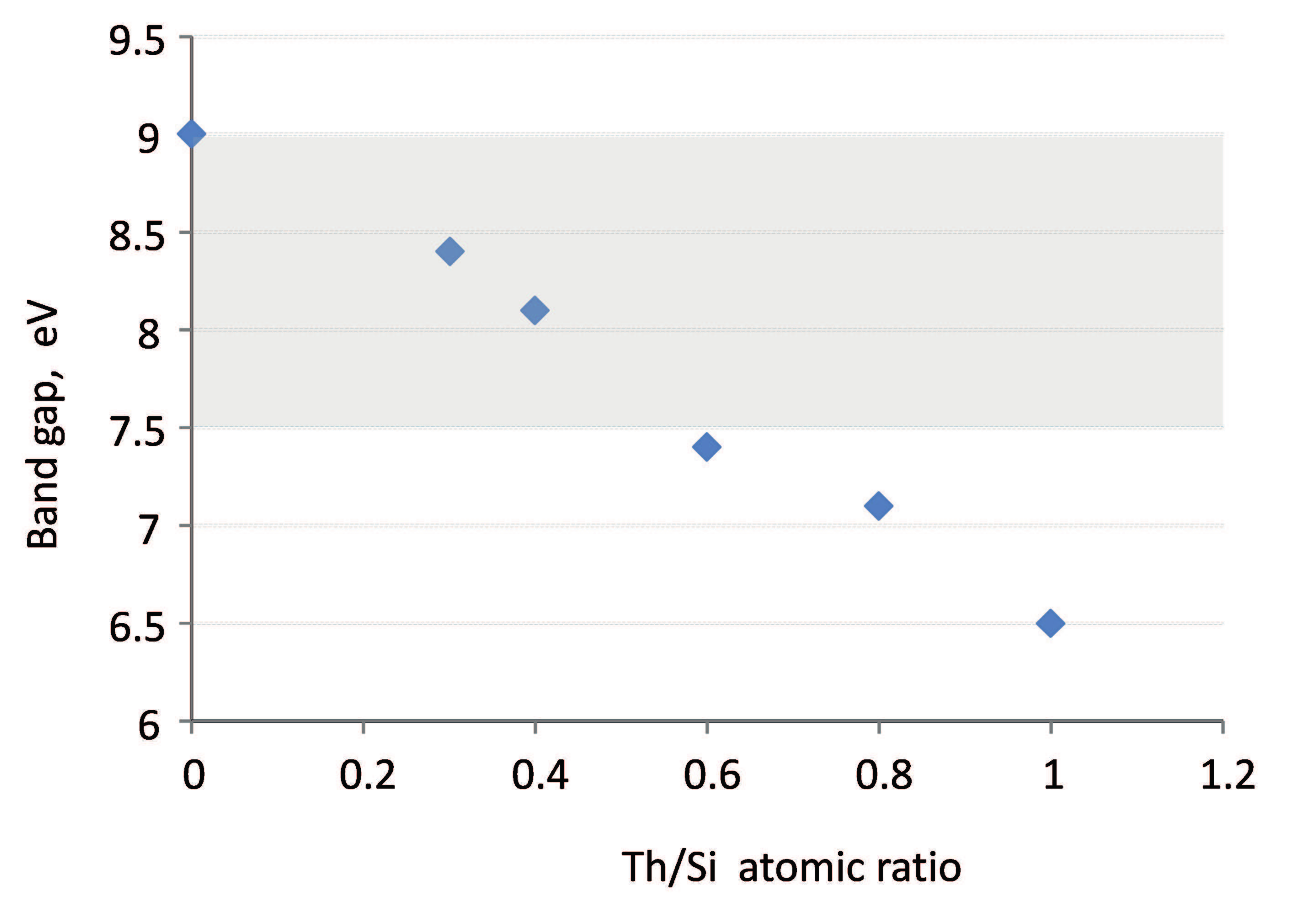}
	\caption{Bandgap depending on local concentration of thorium atoms according  the results of chemical composition analysis of the surface with XPS and REELS methods}
	\label{fig:image4}
\end{figure}

The band gap was restored by the REELS technique realized in the XSAM-800 spectrometer chamber in terms of the procedure described in \cite{24}. Spectra of characteristic electron energy loss with energy 1~keV scattered on the surface of the initial SiO${}_{2}$ film and the same film after Th implantation are shown in Fig.~\ref{fig:image3}. It can be seen that in the process of increasing the amount of implanted thorium the band gap decreases from 9.0 for the original SiO${}_{2}$ film to 6.5~eV for a film with Th/Si~=~1 when the composition becomes close to the compound ThSiO$_4$. It is necessary to note that within the energy range up to 30~eV the structure of the observed REELS spectra is determined not only by the losses of electron energies in the excitation of electron interband transitions but also by the excitation of plasmons. For example, as it follows from the SiO${}_{2}$ spectrum there is the bulk plasmon peak  at the energy of 22~eV that agrees with the data \cite{25}. Existence of two arms in the spectrum of SiO${}_{2}$ at energies of 11~eV and 15~eV can be caused by electron transitions from completed shell corresponding to oxygen 2p-states into unoccupied 3s3p-states of Si \cite{24,26}. {We note that energy-loss spectra observed within the energy range of 1-6 eV demonstrate rather  small but well pronounced  fine structure. The latter  may appear due to the contribution of surface states of non-stoichiometric silicon oxide having the structure. SiO$_x$ ($x<2$) \cite{42} or due to an admixture of carbon \cite{43}. However, it should be underlined these energy losses are small enough and do not noticeably influence the value of useful signal while working with Th-229.} The dependence of the band gap on the thorium concentration in the ThSi${}_{x}$O${}_{y}$ film is shown in Fig.~\ref{fig:image4}.

\section{\textit{Ab initio} calculations}
The multicomponent system consisting of the bulk SiO${}_{2}$ with the secondary phase of ThO${}_{2}$ possesses the properties of a heterogeneous material. In studies of light-matter interactions, if the photon wavelength appears to be smaller than the length of ThO${}_{2}$ interatomic distances, such SiO${}_{2}$-ThO${}_{2}$ composite system can be theoretically considered as the optically homogeneous material. Supposing that the material is formed of N components, the optical properties of the system under consideration can be modelled \cite{27,28} in terms of effective macroscopic dielectric function and refractive index as follows

\begin{equation} \label{GrindEQ__1_} 
\sum _{j=1}^{N}{f}_{j} \frac{{\varepsilon }_{eff}-{\varepsilon }_{j}}{{\varepsilon }_{eff}+{L}_{j} \left( {\varepsilon}_{j}-{\varepsilon}_{eff} \right)} .
\end{equation} 

\noindent Eq.~\eqref{GrindEQ__1_} represents Bruggeman effective medium approximation \cite{28}. The summation over $j$ accounts for the contribution of each component; the parameter ${f}_{j}$ determines the filling factor, and ${L}_{j}$ characterizes the shape of the secondary phase. For spherical form ${L}_{j}$=1/3. Note that, by using in Eq.~\eqref{GrindEQ__1_} spectral distribution of imaginary (${\varepsilon}_{j2}$) and real (${\varepsilon}_{j1}$) parts of the macroscopic dielectric function relating to each moiety, one can model the optical behavior of the composite system depending on the surfactant agent concentration in terms of different ${f}_{j}$. In particular, as is shown below, this approach allowed us to demonstrate the possibility to tune the SiO${}_{2}$ host band gap via incorporation of ThO${}_{2}$ secondary phase by changing the filling factor.

Spectral distributions of the macroscopic dielectric function were theoretically evaluated for ThO${}_{2}$ and SiO${}_{2}$ crystalline materials by ab initio studies based on the density functional theory (DFT). The calculations have been made by using Vienna \textit{ab initio} simulation package (VASP) \cite{29,30,31} together with the projector augmented-wave (PAW) pseudopotential method \cite{5,6,7, 32,33,34}. Exchange and correlation effects have been taken into account within the Perdew-Burke-Ernzerhof (PBE) \cite{35,341} generalized-gradient approximation (f (PBE) GGA) functional \cite{8} or Heyd-Scuseria-Ernzerhofhof  hybrid functional (HSE06) (HSE) \cite{9,10, 36}. The PAW-PBE pseudopotentials employed in calculations correspond to Th(6\textit{s}${}^{2}$6\textit{p}${}^{6}$6\textit{d}${}^{2}$7s${}^{2}$), Si(3\textit{s}${}^{2}$3\textit{p}${}^{2}$), and O(2\textit{s${}^{2}$}2\textit{p${}^{4}$}) valence states. The calculations were performed with a 8x8x8 k-point mesh, and a plane-wave energy cutoff of 600 eV. Structural relaxation was stopped when the total energy change was less than 10${}^{-8 }$~eV, and the residual forces and pressure were less than 10${}^{-4}$~eV/A and 0.06~kB, respectively. 

The optimized (relaxed) lattice parameters of SiO${}_{2}$ and ThO${}_{2}$ have been used for electronic structure calculations within the HSE06 hybrid functional. The band gap estimated as the difference of energy corresponding to the bottommost conduction band and topmost valence band was found to be 5.99 eV for ThO${}_{2}$ and 9.18~eV for SiO${}_{2}$, respectively. Note that these findings are very close to the values of 6.0~eV and 9.0~eV which have been established experimentally in the present work. Next, we performed the single evaluations of the dielectric function for both materials. The obtained ${\varepsilon}_{1}$ and ${\textbf{}}_{2}$ characteristics have been further used as input of Eq.~\eqref{GrindEQ__1_} to derive the frequency profile of the macroscopic dielectric function of the SiO${}_{2}$ - ThO${}_{2}$ composite system in terms of the filling factor. The dependence of the filling factor on ThO${}_{2}$ (considered as the surfactant agent) was expressed by \textit{f}~=~Th/Si. 

\begin{figure}
	\centering
	\includegraphics[width=0.9\linewidth]{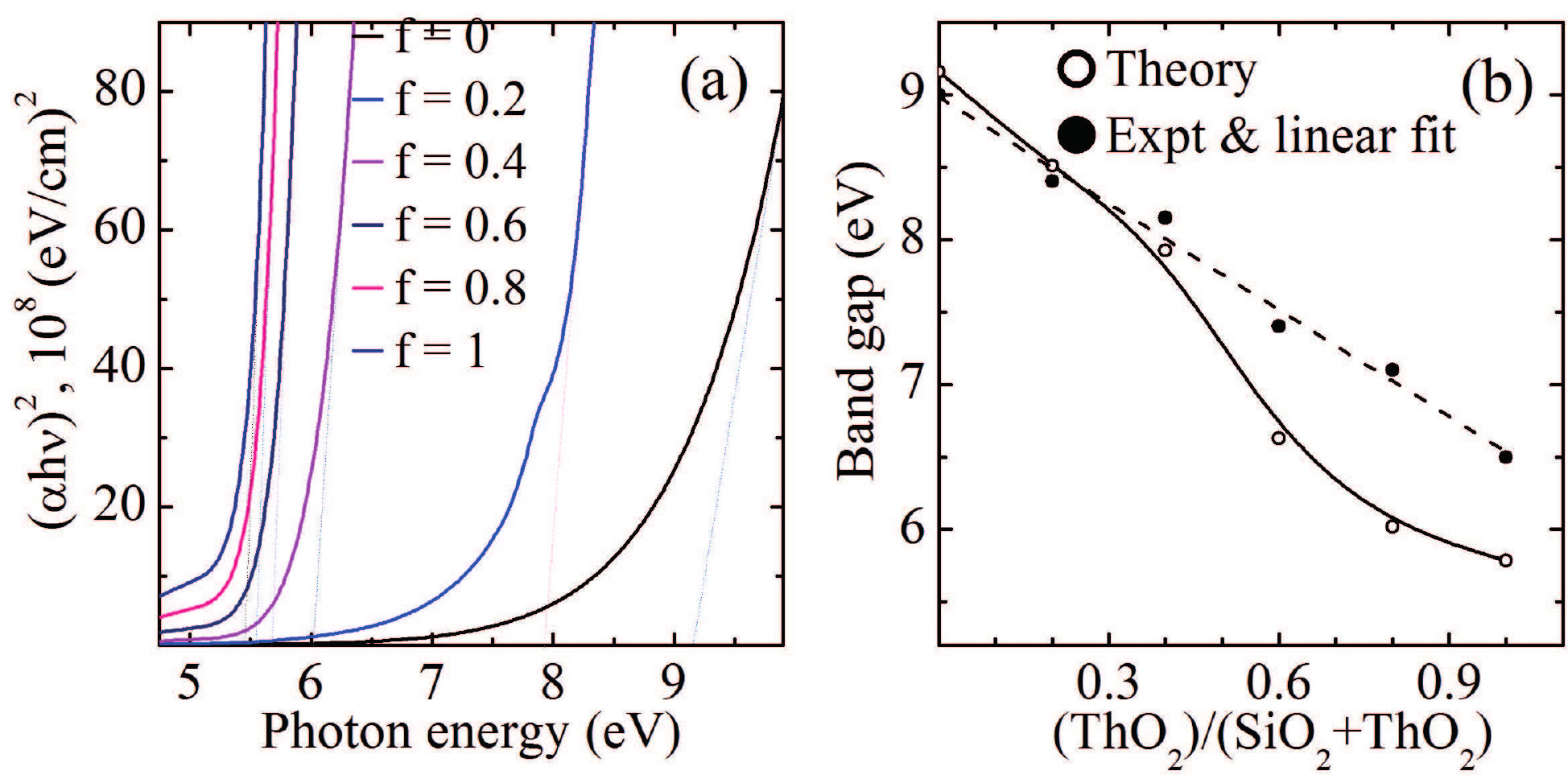}
	\caption{(a) Tauc plot for SiO${}_{2}$ and ThO${}_{2}$ obtained by first-principles calculations and for SiO${}_{2}$/ThO${}_{2}$ composites obtained by Bruggeman effective medium approximation;
	(b) dependence of the band gap on fill factor Th/Si (open circles and solid line) calculated by effective mass theory and (filed circles and dashes) established experimentally}
	\label{fig:image5}
\end{figure}

Absorption spectra $\alpha$($\hbar w$) have been simulated using the calculated ${\varepsilon}_{2}$($w$) and ${\varepsilon}_{1}$($w$). From Tauc plot (see Fig.~\ref{fig:image5}a) band gap for SiO${}_{2}$ and ThO${}_{2}$ as well as for their composites has been estimated. One estimates the Eg from intersection of slope with abscissa. Here $\hbar$ is the Plank constant and w is the frequency of the sunlight. The band gaps estimated from Tauc plot are equal to 5.51~eV for ThO${}_{2}$ and 9.16~eV for SiO${}_{2}$. Figure~\ref{fig:image5}b  shows dependence of the band gap on fill factors of ThO${}_{2}$. Analysis shows that incorporation of the ThO${}_{2}$ causes gradual reduction of the band gap for SiO${}_{2}$. 

The calculated band gaps have been compared to experimental data obtained in this work by REELS. Analysis shows that the dependence of the experimentally measured band gap on fill factor can be approximated with a straight line whereas similar dependence of the theoretically estimated band gap is not linear. Good agreement between theory and experiment is achieved at small fill factors in the range from 0.0~to~0.4. At larger fill factors, the discrepancy increases. The largest difference between the theory and experiment is $\sim$~11\% of the experimentally established band gap. The reason for the discrepancy can be explained by limitations of the effective medium theory. In particular, the fill factor that has been used in this work corresponds to the assumption about non-interacting secondary phases. Furthermore, the secondary phases are assumed to possess spherical shape and homogeneous size. Also, the effective medium theory can be applied to the particular system of ThO${}_{2}$/SiO${}_{2}$ when the ThO${}_{2}$ particle size exceeds 200~nm. {As soon as the total  number of implanted thorium atoms is not large enough we are obviously  allowed to use \textit{Effective medium theory} only for qualitative estimations.} All these limitations contribute on accuracy of the theoretically estimation of the band gaps for the ThO${}_{2}$/SiO${}_{2}$ composite.

\section{Discussion}

According to the experiments and calculations presented in the paper, in the subsurface area a wide band gap compound with a gap width of 9 eV is formed as a result of laser implantation of thorium ions into the SiO${}_{2}$ matrix. These samples are of interest \textit{per~se} for searching for a unique low-lying isomeric state in ${}^{229}$Th by various methods, including synchrotron radiation sources. However the method of laser implantation in this case serves as an exciter of isomeric nuclei of ${}^{229}$Th. In fact, at the early stages of the plasma torch expansion, when plasma can still be treated as quasi-stationary, the ions experience a lot of inelastic collisions with electrons and photons. In addition the most likely process of excitation of nuclei under such conditions is the opposite process of internal electronic conversion IIC \cite{15}. According to this process, plasma electrons populate the ion levels from the states of the continuous spectrum of the energy \textit{E}, i.e., they change into the states of discrete spectrum of the energy $E_{f}$.The nucleus is excited by a virtual photon. This is an inverse to the decay process of the isomeric nuclear state through the gamma ray internal electron conversion channel. The IIC cross-section onto a completely unoccupied ion shell takes the following form
\begin{equation} \label{GrindEQ__2_} 
\sigma _{IIC} (E\to f)\approx \delta (E_{is} -(E-E_{f} ))\frac{\lambda _{e}^{2} }{4} \Gamma _{c} (\omega _{N} ;f)\frac{2J_{F} +1}{2J_{I} +1}  
\end{equation} 
where $\Gamma _{c} (E_{is} ;f)$ is the partial conversion width of the isomeric level upon decay through \textit{ f }-- ion shell, \textit{$\lambda$${}_{e}$} is the electron wavelength with energy $E_{res} $, $J{}_{F,I} $ are spins of isomeric (final) and initial ground states (unit system is $\hbar=c=k=1)$. The cross section $\sigma _{IIC}$ has a pronounced resonance character. Only those electrons of the plasma spectrum "work", or result in the excitation of nuclei whose energy $E_{res} $ coincides with the energy difference of the nuclear transition $\omega _{N}$ and the absolute value of the binding energy $E_{f}$ on the populated atomic shell within the conversion width of the nuclear state $\Gamma$\textit{${}_{c}$}. The IIC process is most effective at the plasma temperature \textit{T} which is comparable to or exceeds the nuclear transition energy $\omega _{N} $\textit{=}7.8~eV several times. Indeed, for these temperature just the states which are important for the internal electronic conversion will be ionised and the number of electrons in the work area of spectrum is relatively high. To achieve such temperatures, it is sufficient to use a laser with an intensity of $I\approx 10^{10}$~ W/cm${}^{2}$ \cite{37}. Now we have a laser offering the pulse energy $E_{L} =100$~mJ under the pulse duration $\tau _{L}~\approx~15$~ns. When focusing into a spot of the diameter of $d=10^{-2}$~cm, we obtain the desired intensity.

With the excitation of nuclei in such a plasma of the electron density $n_{e} $and their energy distribution function $f(E)$, the efficiency of the IIC mechanism can be written as:
\begin{equation} \label{GrindEQ__3_} 
\eta _{IIC} =\int _{0}^{\infty }\sigma _{IIC} (E)\, n_{e} \, f_{e} (E)\, \tau \, \upsilon _{e} \, \frac{dE}{T}   
\end{equation} 
Integration (incorporation) leads to elimination of the energy delta function from $\sigma _{IIC}$. Then in order of magnitude the efficiency of the mechanism proves to be equal to
\begin{equation} \label{GrindEQ__4_} 
\eta _{IIC} \approx \frac{\left(\lambda _{e}^{res} \right)^{2} }{4} \frac{\Gamma _{c}^{tot} }{T} f_{e} (E_{res} )\, n_{e} \, \tau \, \upsilon _{e}^{res} ,  
\end{equation} 
where $\Gamma _{IC}^{tot} $ is the total IC width of the nuclear level ($\Gamma _{IC}^{tot} =\ln 2/7\mu {\rm s}$ \cite{38}), $E_{res} $ is the resonant energy of the electrons determined by the condition (ratio) $E_{res} =E_{is} -|E_{f} |~ \approx~ $1.8~эВ (for 7s electrons the binding energy in the thorium atom is 6.31~eV \cite{39}), $\lambda_{e}^{res}$ is the resonant wavelength of an electron $\lambda_{e}^{res}~=~2\pi /p_{e}^{res} $, $p_{e}^{res}$ is the electron momentum  $p_{e}^{res}~=~m\, \upsilon_{e}^{res}$, аnd its velocity is $\upsilon_{e}^{res}~=~\sqrt{2E_{res}/m}$. Then one can estimate the effective cross section $\sigma _{IIC}^{eff}~\approx~\frac{\left(\lambda_{e}^{res} \right)^{2}}{4} \frac{\Gamma_{c}^{tot}}{T}~=~1.4\times 10^{-26}~$cm${}^{2}$. Taking into account that the lifetime of stationary plasma is $\tau =d/\upsilon \approx 10^{-10} $~s (the estimate of the plasma spread(expansion) time at a temperature of 10 eV), the solid-state density of electrons and thorium nuclei is $n_{Th-229} =3\times 10^{22} $~cm${}^{-3}$, the efficiency of the mechanism for Maxwell's energy distribution$f_{e} (E_{res} )$ reads $\eta _{IIC}~\approx~\sigma _{IIC}^{eff} \, f_{e} (E_{res} )\, n_{e} \, \tau \, \upsilon _{e}^{res} \approx 6.5\times 10^{-6} $. The number of excited nuclei per laser pulse in the interaction region $S=\pi d^{2} /4,$$d=10^{-2} $ cm, $h=10^{-4} $cm (determined by a wavelength of 1.06 $\mu $m) is $N_{is} \approx \eta _{IIC} \; n_{Th-229} \, S\, h\, \approx 10^{9}$. This looks very promising and makes the method of laser implantation absolutely unique, because it allows combining two important processes: 1~--~excitation of the ${}^{229}$Th isomeric nuclei; 2~--~implantation of a wideband dielectric into the matrix. Thus, the estimations made again with experimental results show that the laser ion implantation of thorium ions in SiO${}_{2}$ matrix can be useful for further research of "nuclear clocks" based on low-lying isomeric transition in ${}^{229}$Th isotope with energy 7.8~$\pm$~0.5~eV.

\section*{Acknowledgements}

This research was supported by a grant of Russian Science Foundation (project No 16-12-00001).


\begin{thebibliography}{10}
\expandafter\ifx\csname url\endcsname\relax
  \def\url#1{\texttt{#1}}\fi
\expandafter\ifx\csname urlprefix\endcsname\relax\def\urlprefix{URL }\fi
\expandafter\ifx\csname href\endcsname\relax
  \def\href#1#2{#2} \def\path#1{#1}\fi

\bibitem{1}
W.~G.~Rellergert, S.~T.~Sullivan, D.~DeMille, R.~R.~Greco, M.~P.~Hehlen, R.~A.~Jackson, J.~R.~Torgerson, E.~R.~Hudson, IOP Conf. Ser. Mater. Sci. Eng. 
15 (2010), 012005.

\bibitem{2}
M.~P.~Hehlen, R.~R.~Greco, W.~G.~Rellergert, S.~T.~Sullivan, D.~DeMille, R.~A.~Jackson, E.~R.~Hudson, J.~R.~Torgerson, J.~Lum.
  133 (2013) 91. 

\bibitem{3}
J.~K.~Ellis, Xiao-Dong~Wen, R.~L.~Martin, Inorg.~Chem.
   53~(13) (2014), 6769.

\bibitem{4}
S.~Stellmer, M.~Schreitl, T.~Schumm,  Sci.~Rep. 5 (2015) 15580.

\bibitem{5}
P.~V.~Borisyuk, O.~S.~Vasilyev, Y.~Y.~Lebedinskii, A.~V.~Krasavin, E.~V.~Tkalya, V.~I.~Troyan, R.~F.~Habibulina, E.~V.~Chubunova, V.~P.~Yakovlev, AIP~Advances 6 (2016) 095304.

\bibitem{6}
E.~V.~Tkalya, C.~Schneider, J.~Jeet, E.~R.~Hudson,  Phys.~Rev.~D 92 (2015) 054324.

\bibitem{7}
B.~R.~Beck , J.~A.~Becker, P.~Beiersdorfer , G.~V.~Brown, K.~J.~Moody , J.~B.~Wilhelmy , F.~S.~Porter , C.~A.~Kilbourne , R.~L.~Kelley, Phys.~Rev.~Lett. 98 (2007) 142501.

\bibitem{8}
B.~R.~Beck, C.~Wu, P.~Beiersdorfer, G.~V.~Brown, J.~A.~Becker, K.~J.~Moody, J.~B.~Wilhelmy, F.~S.~Porter, C.~A.~Kilbourne, R.~L.~Kelley, Proceedings of the 12th International Conference on Nuclear Reaction Mechanisms 1 (2010) 255.

\bibitem{9}
E.~Tkalya, Phys.~Rev.~Lett. 106 (2011) 162501.

\bibitem{10}
E.~Peik, M.~Okhapkin, Comptes Rendus Physique 16 (2015) 516-523.

\bibitem{11}
J.~Jeet, C.~Schneider, S.~T. Sullivan, W.~G. Rellergert, S.~Mirzadeh, A.~Cassanho, H.~P.~Jenssen, E.~V. Tkalya, and E.~R. Hudson, Phys.~Rev.~Lett. 114 (2015)  253001.

\bibitem{12}
A.~Yamaguchi, M.~Kolbe, H.~Kaser, T.~Reichel, A.~Gottwald, E.~Peik, New J. Phys. 17 (2015) 053053.

\bibitem{13}
P.~Dessovic, P.~Mohn, R.~A.~Jackson, G.~Winkler, M.~Schreitl, G.~Kazakov. T.~Schumm, J.~Phys.:~Condens.~Matter 26 (2014) 105402.

\bibitem{14}
J.~B.~Amaral, D.~F.~Plant, M.E.G.~Valerio, R.~A.~Jackson, J.~Phys.:~Condens.~Matter 15 (2003) 2523.

\bibitem{15}
V.F.~Strizhov, E.V.~Tkalya, Sov. Phys. JETP 72 (3) (1991) 387.

\bibitem{16}
Y.~Lebedinskii, A.~Zenkevich, Journal of Applied Physics 101 (2007) 074504.

\bibitem{17}
A.A.I. Khalil, Surface and Coatings Technology 200 (2005) 774.

\bibitem{18}
V.~Yu. Fominski, V.~N. Nevolin, I.~Smurov,  J.~Appl.~Phys. 96 (2004) 2374.

\bibitem{39}
J. Chandrappan, M. Murray, T. Kakkar, P. Petrik, E. Agocs, Z. Zolnai, D.P. Steenson, A. Jha, G. Jose, Scientific Reports 5 (2015) 14037 

\bibitem{40}
M.A. Pushkin, V.V. Lebid'ko, V.D. Borman, V.N. Tronin, V.I. Troyan, I. Smurnov, Appl. Surf. Sci., 252 (2006) 4425

\bibitem{19}
Y.A.~Teterin, I.O.~Utkin, I.V.~Melnikov, A.M.~Lebedev, A.Y.~Teterin, K.E.~Ivanov, A.S.~Nikitin, L.~Vukchevich, J.~Struct.~Chem. 41 (2000) 1167.

\bibitem{20}
P.V.~Borisyuk, O.S.~Vasil’ev, V.B.~Loginov, Yu.Yu.~Lebedinskii, V.I.~Troyan, Colloid Journal 76 (5) (2014) 514.

\bibitem{21}
J.F. Ziegler, J. Applied Physics, 85 (1999) 1249.

\bibitem{22}
J.I. Apinaniz, B. Sierra, R. Martinez, A. Longarte, C. Redondo, F. Castano, J.Phys. Chem. C 112 (2008) 16556.

\bibitem{41}
P.V. Borisyuk, S.P. Derevyashkin, K.Yu. Khabarova, N.N. Kolachevsky, Yu.Yu. Lebedinsky, S.S. Poteshin, A.A. Sysoev, E.V. Tkalya, D.O. Tregubov, V.I. Troyan, O.S. Vasiliev, V.P. Yakovlev, V.I. Yudin, European Journal of Mass Spectrometry, 23 (4), (2017) 136

\bibitem{23}
J.M.~Lackner, W.~Waldhauser, R.~Ebner, W.~Lenz, C.~Suess, G.~Jakopic, G.~Leising, H.~Hutter, Surface and Coatings Technology 163 (2003) 300

\bibitem{25}
J.E.~Rowe, Appl.~Phys.~Lett. 25 (1974) 576.

\bibitem{24}
P.V.~Borisyuk, V.I.~Troyan, M.A.~Pushkin, V.D.~Borman, V.N.~Tronin, J.~Nanosci.~Nanotechnol. 12. (2012) 8751.

\bibitem{26}
W.Y.~Ching, Phys.~Rev.~B.  26 (1982) 6633.

\bibitem{42}
A. Barranco, F. Yubero, J.P. Espinos, P. Groening, AR. Gonzalez-Elipe, Journal of Applied Physics 97 (2005) 113714 

\bibitem{43}
D.K. Kim, K.S. Jeong, Y.S. Kang, H.K. Kang, S.W. Cho,  S.O. Kim, D. Suh, S. Kim,  M.H. Cho, Sci Rep. 10 (2016) 34945

\bibitem{27}
A. G. Bruggemann, Ann.~Physik. 24 (1935) 636.

\bibitem{28}
G.A.~Niklasson, C.G.~Granqvist, O.~Hunderi, Appl.~Opt. 20 (1981) 26.

\bibitem{29}
G.~Kresse, J.~Furthmüller, Phys.~Rev.~B.  54 (16) (1996) 11169.

\bibitem{30}
G.~Kresse, J.~Furthmüller, Comput.~Mater.~Sci. 6 (1) (1996) 15.

\bibitem{31}
G.~Kresse, J.~Hafner, Phys. Rev. B. 49 (20) (1994) 14251.

\bibitem{32}
G.~Kresse, D.~Joubert, Phys. Rev. B. 59 (3) (1999) 1758.

\bibitem{33}
G.~Kresse, J.~Hafner, Phys. Rev. B. 47 (1) (1993) 558.

\bibitem{34}
P.E.~Blochl, Phys. Rev. B. 50 (24) (1994) 17953.

\bibitem{35}
J.~Heyd, G.E.~Scuseria, M.~Ernzerhof, J. Chem. Phys. 118 (18) (2003) 8207.

\bibitem{341}
J.P.~Perdew, K.~Burke, M.~Ernzerhof, Phys. Rev. Lett. 77 (18) (1996) 3865.

\bibitem{36}
L.~Torrisi, S.~Gammino, A.~Picciotto, D.~Margarone, L.~Laska, J.~Krasa, K.~Rohlena, J.~Wolowski, Rev. Sci. Instrum. 77 (2006) 03B708

\bibitem{37}
B.~Seiferle, L.~von der Wense, P.G.~Thirolf, Phys. Rev. Lett. 118 (2017) 042501

\bibitem{38}
A.Kramida, Yu. Ralchenko, J. Reader and NIST ASD Team, NIST Atomic Spectra Database (ver. 5.3), [Online]. Available: http://physics.nist.gov/asd. National Institute of Standards and Technology, Gaithersburg, MD.

\end{thebibliography}
\end{document}